\documentclass[conference]{IEEEtran}
\IEEEoverridecommandlockouts

\usepackage{cite}
\usepackage{amsmath,amssymb,amsfonts}
\usepackage{algorithmic}
\usepackage{graphicx}
\usepackage{textcomp}
\usepackage{xcolor}
\usepackage{url}
\usepackage{hyperref}
\usepackage{comment}

\usepackage{tikz}
\usetikzlibrary{shapes.geometric, arrows.meta, positioning}

\tikzset{
  myStep/.style={
    rectangle, rounded corners=4pt,
    minimum width=6.5cm, minimum height=1cm,
    text centered, draw=blue,
    font=\sffamily\footnotesize,
    fill=white!10, text width=7.5cm
  },
  myOptional/.style={
    rectangle, rounded corners=4pt,
    minimum width=6.5cm, minimum height=1cm,
    text centered, draw=blue,
    font=\sffamily\footnotesize\itshape,
    fill=white!20, text width=7.5cm
  },
  myArrow/.style={thick, ->, >=stealth}
}

\def\BibTeX{{\rm B\kern-.05em{\sc i\kern-.025em b}\kern-.08em
    T\kern-.1667em\lower.7ex\hbox{E}\kern-.125emX}}
\begin{document}

\title{DEQSE Quantum IDE Extension: Integrated Tool for Quantum Software Engineering

    \thanks{This paper has been accepted for presentation at the 2025 IEEE International Conference on Quantum Computing and Engineering (QCE).}

    \thanks{This work has been supported by the Research Council of Finland through project DEQSE (349945), and by the Business Finland through project SeQuSoS (112/31/2024).
    
    }
}

\author{\IEEEauthorblockN{Majid Haghparast}
    \IEEEauthorblockA{
    \textit{University of Jyväskylä}\\
    Jyväskylä, Finland \\
    majid.m.haghparast@jyu.fi}
    \and
    \IEEEauthorblockN{Ronja Heikkinen}
    \IEEEauthorblockA{
    \textit{University of Jyväskylä}\\
    Jyväskylä, Finland \\
    ronja.k.heikkinen@jyu.fi}
    \and
    \IEEEauthorblockN{Samuel Ovaskainen}
    \IEEEauthorblockA{
    \textit{University of Jyväskylä}\\
    Jyväskylä, Finland \\
    samuel.s.ovaskainen@student.jyu.fi}
    \and
    \IEEEauthorblockN{Julian Fuchs}
    \IEEEauthorblockA{
    \textit{University of Jyväskylä and University of Potsdam}\\
    Jyväskylä, Finland and Potsdam, Germany\\
    julian.fuchs2@student.hpi.uni-potsdam.de}
    \and
    \IEEEauthorblockN{Jussi P P Jokinen}
    \IEEEauthorblockA{
    \textit{University of Jyväskylä}\\
    Jyväskylä, Finland \\
    jussi.p.p.jokinen@jyu.fi}
    \and
    \IEEEauthorblockN{Tommi Mikkonen}
    \IEEEauthorblockA{
    \textit{University of Jyväskylä}\\
    Jyväskylä, Finland \\
    tommi.j.mikkonen@jyu.fi}
}

\maketitle

\begin{abstract}
    This paper presents a tool that simplifies quantum software development by unifying circuit design, code generation, and execution within a single cross-platform environment that supports iterative development. Implemented as open source, the DEQSE Quantum IDE Extension has been developed to provide quantum functionalities within the Visual Studio Code environment, including project creator, code runner, code converter, and embedded quantum circuit simulator. Furthermore, the system provides capabilities that facilitate iterative development and support learning, distinguishing it from other available Visual Studio Code Extensions for quantum computing.
\end{abstract}


\section{Introduction} \label{section:introduction}
    
    The growing interest in quantum computing has increased the demand for reliable quantum software. However, quantum programming poses unique challenges and requires tools adapted to quantum mechanics. Current solutions often depend on specific platforms and vendors, limiting accessibility and creating lock-in~\cite{Murillo:2024, Scekic:2022, Matthews:2021}. Prior studies highlight the need for specialized tools and methodologies to support quantum software development~\cite{Khan:2024b,dwivedi2024quantum}. 
    
    To address these challenges, we propose the DEQSE\footnote{Developer Experience in Iterative Quantum Software Engineering} Quantum IDE Extension\footnote{\url{https://github.com/DEQSE-Project/deqse-vscode-extension}}$^{,}$\footnote{\url{https://marketplace.visualstudio.com/items?itemName=JYUQICTeam.deqse}} (shortened to DEQSE Extension hereafter) for Visual Studio Code. This extension provides features such as code converter, an embedded quantum circuit simulator, project creator and code runner, to streamline quantum programming tasks. The embedded quantum circuit simulator, Quirk-E\footnote{\url{https://quirk-e.dev}}$^{,}$\footnote{\url{https://github.com/DEQSE-Project/Quirk-E}}, is an extended version of Quirk\footnote{\url{https://algassert.com/quirk}}$^{,}$\footnote{\url{https://github.com/Strilanc/Quirk}}. It allows developers to work with quantum circuits in its drag-and-drop interface.

    Approaches like this contribute to improving workflow through the reduction of system complexity and the optimization of workflow efficiency, a goal that our DEQSE Extension shares as we facilitate experimentation with quantum computing in quantum software development. Based on our evaluation, the proposed DEQSE Extension provides functionalities in a way which differentiates it from other tools and highlights its value for cross-framework development and learning by combining a visual circuit designer with code generation and execution, cross-platform compatibility, and support for iterative development workflows.
    
    The rest of this paper is structured as follows: Section~\ref{section:background} outlines the challenges in quantum computing that motivate our work. Section~\ref{section:implementation} describes the technical details and functionalities of the proposed DEQSE Extension. Section~\ref{section:evaluation} represents a demonstrative use case scenario and reviews related solutions, highlighting the advantages of our DEQSE Extension. Section~\ref{section:discussion} discusses the results, identifies threats to validity, and suggests directions for future research. Finally, Section~\ref{section:conclusions} concludes the paper.
    
\section{Background and Motivation} \label{section:background} 
    
    Fundamental differences between classical and quantum computing, including concepts such as qubits, superposition, interference, entanglement, and teleportation render many conventional tools insufficient~\cite{Jones-EtAl:2019, Sanchez-and-Alonso:2021,babu2025gate}. Classical bits are either 0 or 1, whereas quantum bits (qubits) can exist in a superposition of both states simultaneously. The sum of the squared magnitudes of the complex amplitudes equals one~\cite{DaSilvaFeitosa-EtAl:2019}. Moreover, classical computing is deterministic in its nature, while quantum computing relies on probability and statistics~\cite{Khan:2024b, Sanchez-and-Alonso:2021}. Quantum computers are much more sensitive to noise than classical computers, which makes error correction and precise gate-level control more critical. These fundamental differences mean that quantum software development requires specialized tools that can represent and manage probabilistic behavior, visualize complex quantum states and circuits, provide low-level gate manipulation when needed, and support robust testing and simulation to mitigate errors and noise. 
    
    The requirement to work at low levels of abstraction, typically at the gate level, due to the lack of widely accepted higher-level abstractions~\cite{Chong-EtAl:2017} significantly increases the complexity of quantum software engineering. This challenge highlights the need for tools that provide more intuitive representations, such as source code and circuit diagrams, and support seamless switching between them.
    
    Iterative development is a key practice in classical software engineering and remains feasible for quantum software engineering during early-stage simulation, because in later stages large-scale quantum executions are computationally expensive, time-consuming, and often non-repeatable~\cite{khan2022embracing}. Additionally, vendor lock-in remains a significant barrier, since many platforms are tightly coupled with specific hardware providers and are incompatible with alternative frameworks~\cite{Murillo:2024, Matthews:2021}. This fragmentation limits developers’ flexibility and discourages widespread adoption.
    
    The above challenges emphasize the urgent need for specialized tools that support higher-level abstractions, cross-platform compatibility, and iterative development workflows~\cite{Jones-EtAl:2019, Murillo:2024, Sepulveda-EtAl:2025}. Integrated environments that unify circuit simulation, testing, coding and execution within a single interface can help mitigate complexity and enhance the developer experience. Developing such tools is critical for advancing quantum software engineering and making quantum computing more accessible.
    
    To meet the above challenges, we derived the following requirements that guided the design of the DEQSE Extension:
    \begin{enumerate}
        \item \textbf{Intuitive dual representation:} The extension must provide both a visual quantum circuit designer and an equivalent source code view, with seamless synchronization between them.
        \item \textbf{Cross-platform compatibility:} The extension should support multiple hardware backends and avoid vendor lock-in by generating portable, framework-independent code.
        \item \textbf{Integrated simulation and testing:} The extension must enable developers to simulate and test quantum circuits easily within the same environment to support iterative workflows.
        \item \textbf{Low-level control when needed:} While offering higher-level abstractions, the extension should allow developers to fine-tune gate-level operations if necessary.
        \item \textbf{User-friendly interface:} The extension should focus on usability to lower the entry barrier for both beginners and experienced quantum developers.
    \end{enumerate}
    
\section{Implementation} \label{section:implementation}
    
    \subsection{Software architecture and technical implementation details} \label{subsection:architecture}
        
        Visual Studio Code\footnote{\url{https://code.visualstudio.com/}} was chosen as an IDE to build the DEQSE Extension. We utilize GitHub\footnote{\url{https://github.com/}} for version control and collaborative code management. Regarding the choice of programming languages, the DEQSE Extension is written in TypeScript, and uses the JavaScript Quantum Circuit Library\footnote{\url{https://github.com/quantastica/quantum-circuit}}$^{,}$\footnote{\url{https://quantastica.com}}$^{,}$\footnote{ \url{https://www.npmjs.com/package/quantum-circuit}}. For other tools and services, the DEQSE Extension uses Yeoman\footnote{\url{https://yeoman.io}} 
        and the Yeoman generator code\footnote{\url{https://www.npmjs.com/package/generator-code}}. The dependencies for the DEQSE Extension include Codicons\footnote{\url{https://github.com/microsoft/vscode-codicons}}, WebView UI Toolkit\footnote{\url{https://github.com/microsoft/vscode-webview-ui-toolkit}} and Visual Studio Code project templates\footnote{\url{https://github.com/cantonios/vscode-project-templates}}. npm is used as a package manager and Font Awesome\footnote{\url{https://github.com/FortAwesome/Font-Awesome}} is used for icons. 
        
\begin{figure}
            \centering
            \includegraphics[width=0.99\columnwidth]{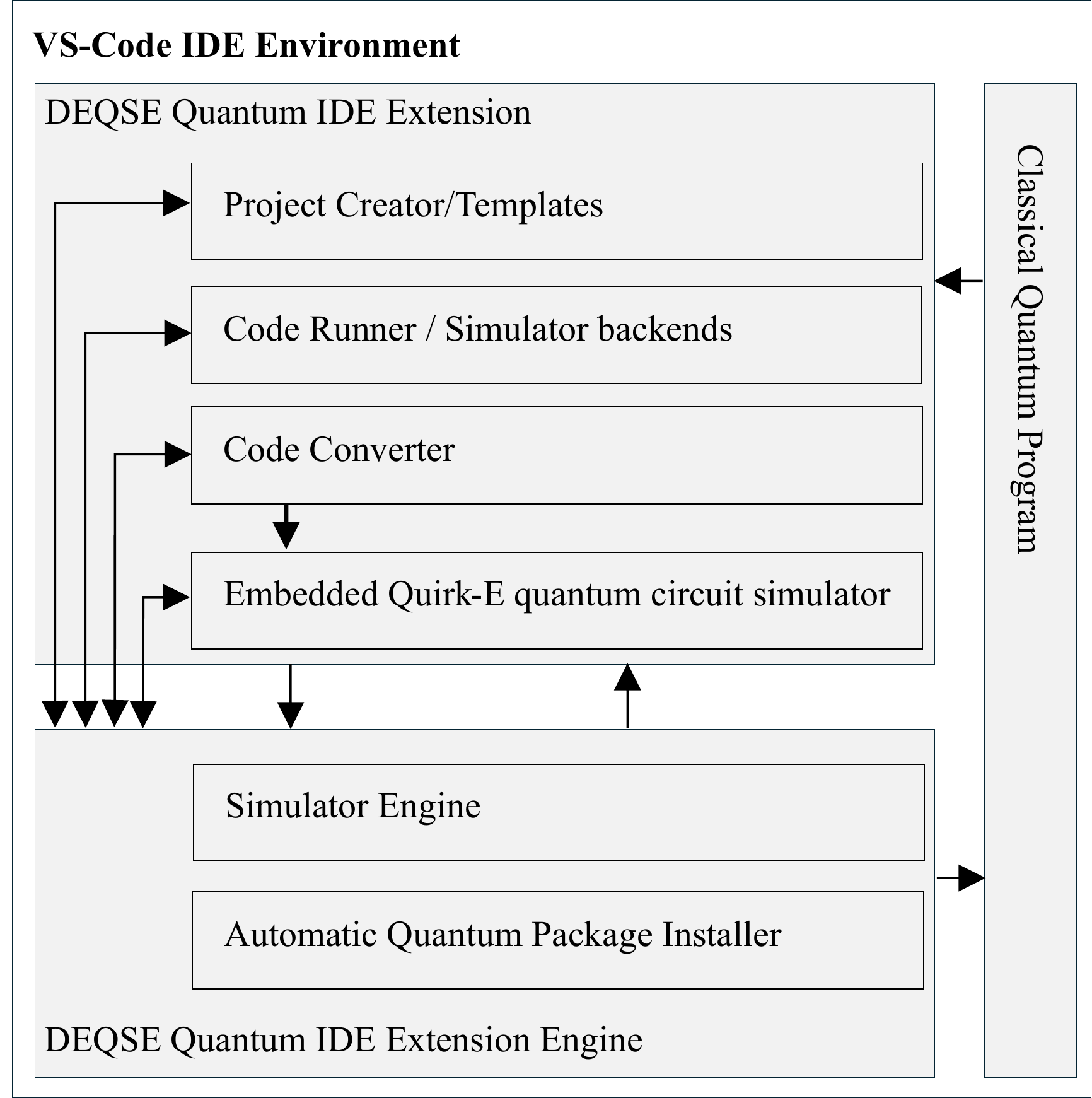}
            \caption{Architecture overview of DEQSE Extension.}
            \label{figure:architecture}
        \end{figure}

        Figure~\ref{figure:architecture} depicts the architecture of the DEQSE Extension. The extension is implemented for Visual Studio Code as part of its architecture. Among the implemented functionalities, the embedded quantum circuit simulator constitutes a key component of this architecture, enabling users to simulate quantum programs directly within the development environment.

    \subsection{Software functionalities description} \label{subsection:functionalities}
        
        The current version of the extension (DEQSE v0.1.0) implements four different key functionalities that are provided in separate menus. Where applicable and feasible, those tools are interconnected with each other in a way that further improves the user experience: the embedded quantum circuit simulator is connected to the code runner. The output of the code runner can be opened in the embedded circuit designer.

        \textbf{Project Creator.} The project creator streamlines project setups by enabling users to initialize new projects from predefined or custom templates. This speeds up configuration, offers language-specific guidance, and reduces the entry barrier for beginners. A predefined Qiskit\footnote{\url{https://www.ibm.com/quantum/qiskit}}$^{,}$\footnote{\url{https://github.com/Qiskit}} template (qiskit-hello-world) is provided, and users can also define their own templates.
        
\begin{figure}
            \centering
            \includegraphics[width=0.99\columnwidth]{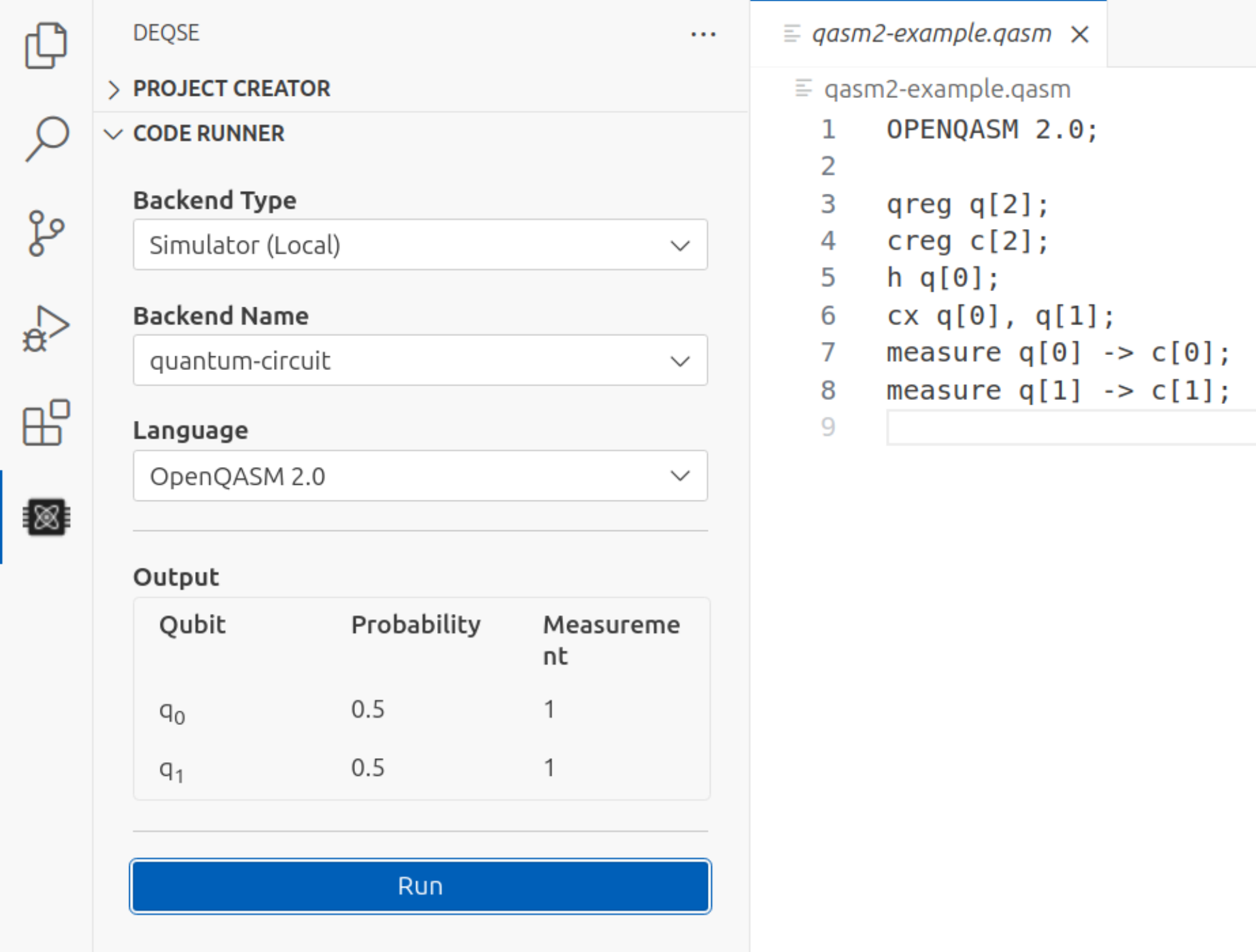}
            \caption{Menu of the extension for the Code Runner functionality. Users can execute their programs within the DEQSE Extension and view the output probabilities.}
            \label{figure:deqse-code-runner}
        \end{figure}

        \textbf{Code Runner.} The code runner feature enables the execution of quantum code directly within the extension environment, providing real-time feedback and simplifying the testing and debugging processes for quantum developers. The code runner supports OpenQASM 2.0, IonQ, Qobj, Qubit Toaster, Quil 2.0 and quantum-circuit as languages for the executable code. It runs the code of the currently opened and focused file window as seen in Figure~\ref{figure:deqse-code-runner} and returns the measurement probabilities. In the backend, a built-in local simulator is available. Additionally, two other simulator options are supported: one through the included JavaScript library Quantum Circuit, and another through the optional proprietary Quantastica Qubit Toaster\footnote{\url{https://pypi.org/project/quantastica-qiskit-toaster/}}$^{,}$\footnote{\url{https://quantastica.com/toaster/}}.
        
\begin{figure}
            \centering
            \includegraphics[width=0.99\columnwidth]{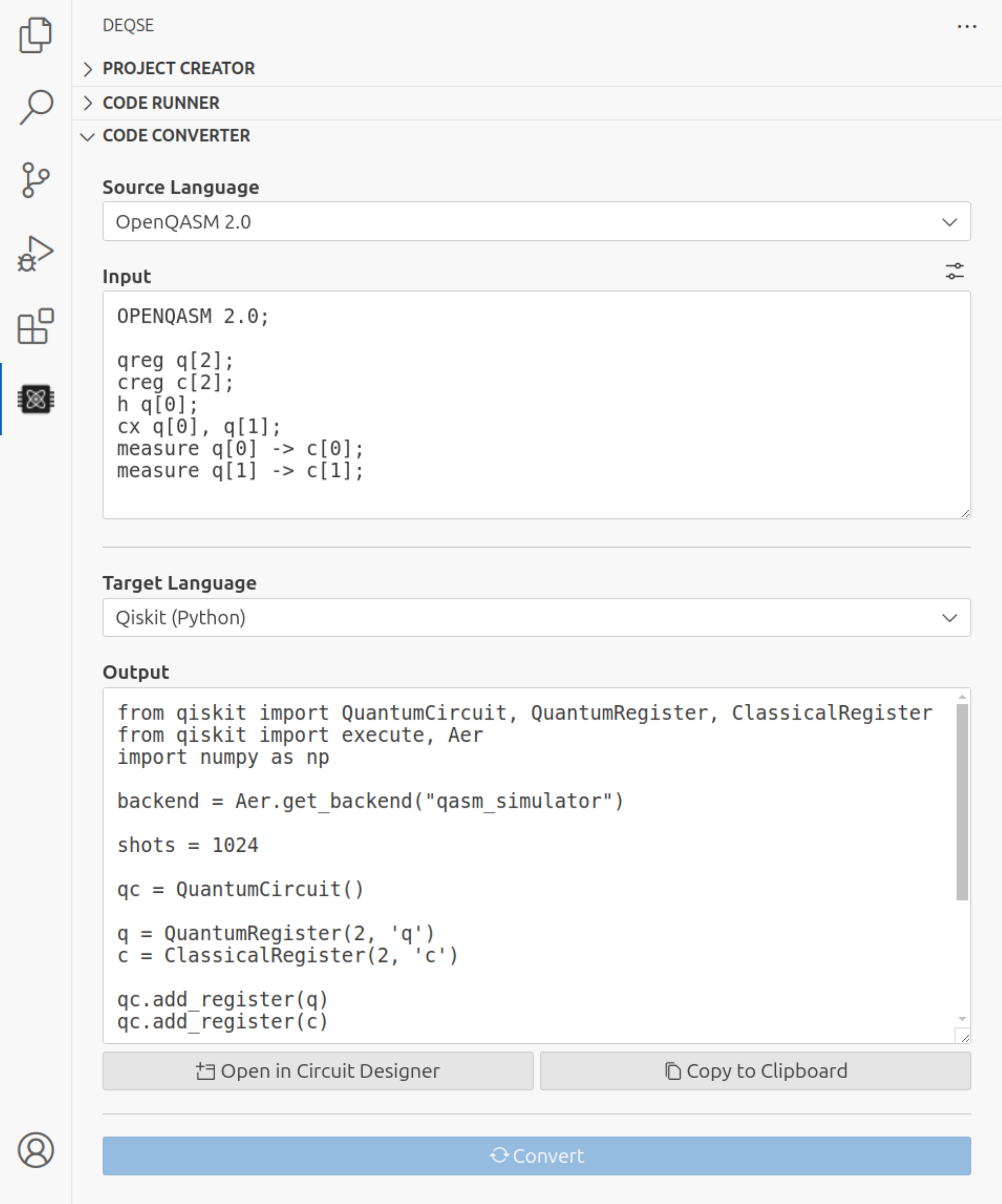}
            \caption{Code converter part of the DEQSE Extension, converting code from OpenQASM 2.0 to Qiskit, as an example.}
            \label{figure:deqse-code-converter}
        \end{figure}

        \textbf{Code Converter.} The code converter (Figure~\ref{figure:deqse-code-converter}) enables users to convert quantum code to different quantum programming languages. This functionality is made using external libraries, explained in section~\ref{subsection:architecture}. The source languages are IonQ, OpenQASM, Qobj, quantum-circuit, Qubit Toaster, and Quil 2.0, whereas the target languages are AQASM, Amazon Bracket, Cirq, IonQ, OpenQASM 2.0, Q Sharp, Qiskit, quantum-circuit, Qubit Toaster, QuEST, Quil, Quirk, TensorFlow Quantum, pyAQASM, and pyQuil.
        
\begin{figure*}
            \centering
            \includegraphics[width=2\columnwidth]{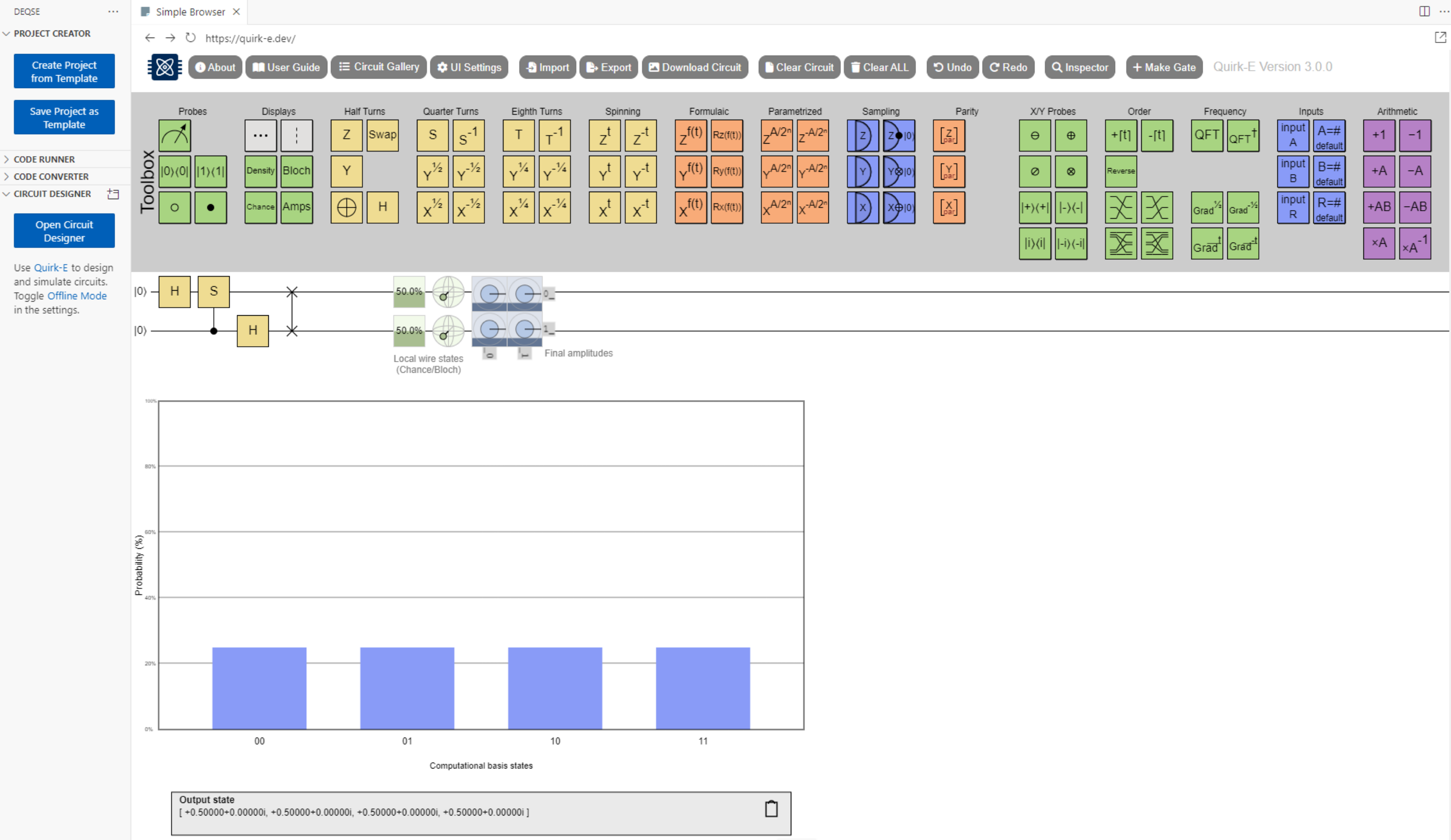}
            \caption{The DEQSE Extension has an embedded circuit designer. The Figure shows DEQSE Extension menu and 2 qubit Quantum Fourier Transform.}
            \label{figure:deqse-circuit-simulator}
        \end{figure*}

        \textbf{Circuit Designer.} The circuit designer feature, embedded quantum circuit simulator (Figure~\ref{figure:deqse-circuit-simulator}), enables developers to simulate their circuits or test their algorithms directly within the DEQSE Extension. Additionally, the extension has the ability to connect directly to the online version of the embedded quantum circuit simulator. While Quirk-E quantum circuit simulator is embedded within the DEQSE extension for demonstration purposes, the DEQSE Extension is designed to be modular, allowing any quantum circuit simulator—such as Quirk or others with similar capabilities—to be embedded.
        
\section{Evaluation and Comparisons} \label{section:evaluation}
    
    \subsection{Demonstrative use case}
        
\begin{figure}[h]
            \centering
            \begin{tikzpicture}[node distance=1.3cm]
                \node (s1) [myStep] {\textbf{Create a new project}, either from scratch or using a predefined template.};
                \node (s2) [myStep, below of=s1] {\textbf{Write QASM code}, e.\,g. Bell state.};
                \node (s3) [myStep, below of=s2] {\textbf{Execute the code} using the ``Run'' button in the ``Code Runner'' to observe the resulting probability distributions.};
                \node (s4) [myStep, below of=s3] {\textbf{Convert the code} from QASM to Qiskit (or other languages) using the ``Code Converter''.};
                \node (s5) [myStep, below of=s4] {\textbf{Visualize the circuit} using the ``Open in Circuit Designer`` in the ``Code Converter''.};
                \node (s6) [myOptional, below of=s5] {\textit{(Optional)} \textbf{Save the project as a reusable template} for future use.};
                
                \draw [myArrow] (s1) -- (s2);
                \draw [myArrow] (s2) -- (s3);
                \draw [myArrow] (s3) -- (s4);
                \draw [myArrow] (s4) -- (s5);
                \draw [myArrow] (s5) -- (s6);
            \end{tikzpicture}
            \caption{Example user flow for cross-framework prototyping with the DEQSE Extension}
            \label{fig:deqse-flow}
        \end{figure}
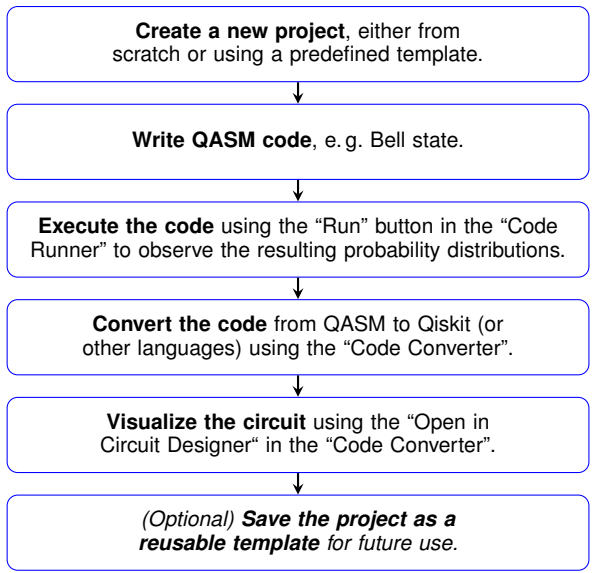

        In response to the distinct challenges of quantum software development -- such as low-level abstraction~\cite{Chong-EtAl:2017}, probabilistic behavior~\cite{Khan:2024b, Sanchez-and-Alonso:2021, DaSilvaFeitosa-EtAl:2019}, and fragmented tooling~\cite{Jones-EtAl:2019, Murillo:2024, Sepulveda-EtAl:2025} -- the DEQSE Extension supports an integrated and iterative development flow for quantum computing. While many existing tools require users to manually switch between environments for writing, visualizing, converting, and executing quantum code, our DEQSE Extension supports these tasks within a unified interface. This might not only improve efficiency but also reduce cognitive load, particularly during early-stage development and experimentation. The following scenario, shown in Figure~\ref{fig:deqse-flow}, illustrates how these capabilities can support a more accessible and reproducible quantum development process. In this scenario, a researcher who routinely works with OpenQASM seeks to streamline their workflow by creating a reusable project setup. Given that their daily tasks often involve similar code structures and project configurations, they aim to establish a consistent starting point for future work. Additionally, the researcher wishes to leverage their existing QASM-based programs to explore tools and platforms built on the popular Qiskit framework, facilitating cross-framework experimentation and easing the transition between language ecosystems.
        
        The process begins by creating a new project, either from scratch or by selecting a predefined/saved template provided by the DEQSE Extension. This is particularly useful for researchers who frequently begin from the same setup. The user then writes OpenQASM code. Unlike standard Visual Studio Code environment, which does not support execution of QASM code from file out-of-the-box, this extension enables in-place execution via the Code Runner feature, allowing immediate observation of probability distributions. This part of the use case is illustrated in Figure~\ref{figure:deqse-code-runner}.
        
        Also, the code can be converted into other languages, e.\,g. Qiskit, a popular quantum computing framework, facilitating further development. This feature is shown in Figure~\ref{figure:deqse-code-converter}. At any point, the circuit can also be visualized using the embedded Circuit Designer for inspection, testing or documentation. Once the user is satisfied with the configuration and functionality, the project can optionally be saved as a reusable template. This feature supports reproducible workflows and reduces setup overhead for iterative experiments or educational settings.
        
    \subsection{Evaluation methods and comparable solutions}
        
        This evaluation examines how the proposed DEQSE Extension supports the quantum software development workflow relative to similar tools. We analyze the functionalities offered by three Visual Studio Code extensions for quantum computing. These extensions are identified through an internet search. Extensions designed exclusively for specific SDKs and platforms, such as those for Aqora, Azure or Intel, were excluded. Although some of the selected extensions lack fully integrated user interfaces in Visual Studio Code, they share a similar general purpose.
        
        QBraid's Quantum Console\footnote{\url{https://marketplace.visualstudio.com/items?itemName=qBraid.quantum-console}} provides integration with QBraid's platform, focusing on cloud service management and job monitoring. UC Quantum Lab Extension\footnote{\url{https://marketplace.visualstudio.com/items?itemName=UCQuantumLab.uc-quantum-lab}}$^{,}$\footnote{\url{https://github.com/UC-Advanced-Research-Computing/UC-Quantum-Lab}} is an educational extension based on Qiskit. Furthermore, Q extension by QuLearnLabs\footnote{\url{https://marketplace.visualstudio.com/items?itemName=LurnDigital.Q-by-QuLearnLabs}}$^{,}$\footnote{\url{https://github.com/QuLearnLabs/Q}} focuses on learning quantum computing, especially for Qiskit. 
    
    \begin{table*}[t]
        \centering
        \caption{Comparison of extensions' functionalities}
        \resizebox{\linewidth}{!}{%
            \begin{tabular}{p{1.5cm}p{1.1cm}p{1.7cm}p{1.5cm}p{1.5cm}p{1.3cm}p{1.3cm}p{1.5cm}p{1.1cm}p{1.1cm}p{1.5cm}p{1.3cm}p{1.3cm}p{1.3cm}}
                \hline
                \textbf{Extension} & \textbf{Lang. Conv.} & \textbf{Code Editor Support} & \textbf{Circuit Designer} & \textbf{Local Simulator} & \textbf{Cloud QPU} & \textbf{Debugging} & \textbf{Result Visuals} & \textbf{VS Code Icon} & \textbf{UI menu sidebar} & \textbf{Templates} & \textbf{AI Assistant} & \textbf{Open Source} & \textbf{UI Views} \\
                \hline
                DEQSE & Yes & Yes & Yes & Yes & No & Yes & Yes & Yes & Yes & Yes & No & Yes & Yes \\
                Q (QuLearnLabs) & No & Yes & Yes & Yes & No & No & No & Yes & Yes & Yes & Yes & Yes & Yes\\
                Q. Console (qBraid) & No & No & No & Yes & Yes & No & Yes & Yes & Yes & No & No & No & No\\
                UC Q. Lab & No & Yes & Yes & Yes & No & No & Yes & No & No & Yes & No & Yes & Yes \\
                \hline
            \end{tabular}
        }
        \label{table:functionalities-comparison}
    \end{table*}

    \begin{table}[ht!]
        \centering
        \caption{Use case comparison}
        \resizebox{\columnwidth}{!}{
            \begin{tabular}{p{5cm}p{5cm}}
                \hline
                \textbf{Use Case} & \textbf{Recommended Extensions} \\
                \hline
                Cross-Framework Prototyping & DEQSE 
                \\
                Cloud Job Management (Multi-backend) & Quantum Console 
                \\
                Qiskit Visualization \& Learning & Q, UC Quantum Lab \\
                AI-Powered Quantum Learning & Q \\
                \hline
            \end{tabular}
        }
        \label{table:use-case-recs}
    \end{table}
    
    \subsection{Evaluation results}
        
        The comparison of functionalities is presented in \autoref{table:functionalities-comparison}, with further explanation and discussion provided in this section.
        Two of the chosen extensions, UC Quantum Lab and Q, target mainly individual quantum frameworks. In contrast, DEQSE Extension appears to be the best for cross-framework conversion, supporting five input languages and additional export languages.
        
        Qiskit visualization is best supported by Q and UC Quantum Lab, both offering predominantly view-only visualizers for code. Quantum Console lacks integrated code editing capabilities since it focuses on cloud monitoring. The DEQSE Extension distinguishes itself by offering an embedded circuit designer tightly integrated with its code conversion feature. The embedded circuit designer includes valuable tools such as an inspector, predefined examples, import, export, and probability histogram. Quantum Console provides the most comprehensive cloud job management across backends. Local simulation is available in DEQSE Extension, UC Quantum Lab, and Q. Quantum Console offer robust support for local and cloud-based simulation. Most extensions include user interface panels and views, and the majority also feature dedicated icons within the Visual Studio Code sidebar and menus located on the left side of the interface.
        
        \autoref{table:use-case-recs} compares use case suitability. Each extension offers distinct advantages, with DEQSE Extension providing effective support for circuit design, education, and cross-framework conversion.
        
\section{Discussion}\label{section:discussion}
    
    The DEQSE Extension offers a set of functionalities, including project creator, code runner, code converter, and an embedded circuit designer. While other Visual Studio Code extensions for quantum computing exist, DEQSE Extension shows comparative advantages in cross-framework prototyping (see Table~\ref{table:functionalities-comparison} and Table~\ref{table:use-case-recs}). Moreover, by integrating essential tools and abstractions within a single environment, the proposed DEQSE Extension facilitates early-stage iterative development, addressing a key challenge identified in quantum software engineering practice.
    
    In the following, we revisit the key requirements defined for DEQSE Extension and discuss how they were addressed in the implementation:
        \begin{enumerate}
            \item \textbf{Intuitive dual representation:} The DEQSE Extension has both an embedded quantum circuit designer and code editor. These two are connected together with the code runner functionality. This way, the user can see both the circuit and the code, and modify either of these inside the IDE.
            \item \textbf{Cross-platform compatibility:} The DEQSE Extension offers backend simulators but can be enhanced with real quantum computer connections. We provide code conversion between multiple languages within the DEQSE Extension. Additionally, code runner currently supports five different languages.
            \item \textbf{Integrated simulation and testing:} The DEQSE Extension enables interactive circuit design and inspector-based debugging of circuits directly within the IDE. The designed circuits can be run in the extension and turned into corresponding code. Multiple programming languages are supported at present. 
            \item \textbf{Low-level control when needed:} Users can write or modify code in the DEQSE Extension. They can also modify circuits and create custom circuits in the embedded circuit simulator.
            \item \textbf{User-friendly interface:} DEQSE Extension provides user interface panels in Visual Studio Code for the functionalities. The user interface panels in the extension are grouped based on the functionalities and can be opened and closed to fit the needs of the user. Templates provide ease of use for quantum software developers. 
        \end{enumerate}
    
    \textbf{Threats to Validity}. 
        The potential threats to the validity of this work are discussed following the classification of Wohlin and others~\cite{Wohlin-EtAl:2012}. The evaluation focused on comparing functionalities and making qualitative judgments about usability and functionality. However, without standardized metrics or user studies, the identified advantages may not accurately reflect user experience or practical effectiveness, posing a threat to construct validity.
        
        As this is the initial version of the DEQSE Extension, it may still include bugs, missing features, or unintended behavior. The demonstration scenario, selected by the authors, could unintentionally highlight strengths while overlooking limitations. Since the evaluation was conducted solely by the developers, the results may be subject to confirmation bias, affecting internal validity. These factors also limit generalizability, impacting external validity. Finally, while access to the tool and source code supports reproducibility, the lack of full automation and independent replication may affect reliability.
        
    \textbf{Future Research}. 
        Future research could focus on improving the DEQSE Extension and developing more functionalities for it, which would benefit the quantum computing community. We will continuously work on the development of DEQSE Extension, for example focusing on connecting DEQSE Extension to real quantum computers, as well as adding quantum algorithms to the DEQSE Extension. The DEQSE extension is open-source, enabling future development and contributions from the broader community. Furthermore, the impact of the DEQSE Extension on learning quantum computing and quantum programming~\cite{haghparast2024innovative} and streamlining quantum software development processes should be further researched. Regarding the perceived value of the DEQSE Extension, its impact on developers, for example in their cognitive load and work efficiency, could be researched.

\section{Conclusions} \label{section:conclusions}
    
    This paper contributes to quantum software development by enabling a more interactive development process and facilitating agile, iterative workflows. We package different quantum software development functionalities to a Visual Studio Code extension in order to streamline the quantum software development workflow. Furthermore, this extension facilitates faster adoption of quantum programming, thus aiding developers' transition to the quantum computing field. The DEQSE Extension can be used in education and research for quantum computing related topics. 
    
    As a conclusion from the evaluation, DEQSE Extension supports cross-framework development by integrating a circuit designer, enabling code conversion, and linking the circuit design environment with the code editor within the IDE.
    The proposed approach fills a significant gap in the availability of practical quantum development tools, especially in the context of Visual Studio Code. The DEQSE Extension will be available permanently on GitHub as an open-source project. The development of such tools contributes to the wider adoption of quantum computing in academic and industry settings.

\bibliographystyle{IEEEtran}
\bibliography{bibliography}

\begin{thebibliography}{10}
\providecommand{\url}[1]{#1}
\csname url@samestyle\endcsname
\providecommand{\newblock}{\relax}
\providecommand{\bibinfo}[2]{#2}
\providecommand{\BIBentrySTDinterwordspacing}{\spaceskip=0pt\relax}
\providecommand{\BIBentryALTinterwordstretchfactor}{4}
\providecommand{\BIBentryALTinterwordspacing}{\spaceskip=\fontdimen2\font plus
\BIBentryALTinterwordstretchfactor\fontdimen3\font minus \fontdimen4\font\relax}
\providecommand{\BIBforeignlanguage}[2]{{%
\expandafter\ifx\csname l@#1\endcsname\relax
\typeout{** WARNING: IEEEtran.bst: No hyphenation pattern has been}%
\typeout{** loaded for the language `#1'. Using the pattern for}%
\typeout{** the default language instead.}%
\else
\language=\csname l@#1\endcsname
\fi
#2}}
\providecommand{\BIBdecl}{\relax}
\BIBdecl

\bibitem{Murillo:2024}
J.~M. Murillo, J.~Garcia-Alonso, E.~Moguel, J.~Barzen, F.~Leymann, S.~Ali, T.~Yue, P.~Arcaini, R.~P{\'e}rez-Castillo, I.~G.~R. de~Guzm{\'a}n \emph{et~al.}, ``Challenges of quantum software engineering for the next decade: The road ahead,'' \emph{CoRR}, 2024.

\bibitem{Scekic:2022}
M.~\v{S}{\'c}eki{\'c} and A.~Yakaryilmaz, ``Comparing quantum software development kits for introductory level education,'' \emph{Baltic Journal of Modern Computing}, vol.~10, no.~1, pp. 87--104, 2022.

\bibitem{Matthews:2021}
D.~Matthews, ``How to get started in quantum computing,'' \emph{Nature}, vol. 591, no. 7848, pp. 166--167, 2021.

\bibitem{Khan:2024b}
A.~A. Khan, D.~Taibi, C.~M. Perrault, and A.~A. Khan, ``Advancing quantum software engineering: A vision of hybrid full-stack iterative model,'' 2024.

\bibitem{dwivedi2024quantum}
K.~Dwivedi, M.~Haghparast, and T.~Mikkonen, ``Quantum software engineering and quantum software development lifecycle: a survey,'' \emph{Cluster Computing}, vol.~27, no.~6, pp. 7127--7145, 2024.

\bibitem{Jones-EtAl:2019}
T.~Jones, A.~Brown, I.~Bush, and S.~C. Benjamin, ``Quest and high performance simulation of quantum computers,'' \emph{Scientific Reports}, vol.~9, no.~1, pp. 1--11, 2019.

\bibitem{Sanchez-and-Alonso:2021}
P.~Sánchez and D.~Alonso, ``On the definition of quantum programming modules,'' \emph{Applied Sciences}, vol.~11, no.~13, p. 5843, 2021.

\bibitem{babu2025gate}
A.~P. Babu, O.~Kerppo, A.~Mu{\~n}oz-Moller, M.~Haghparast, and M.~Silveri, ``Gate teleportation-assisted routing for quantum algorithms,'' \emph{Quantum Science and Technology}, vol.~10, no.~3, p. 035004, 2025.

\bibitem{DaSilvaFeitosa-EtAl:2019}
S.~d.~S. Feitosa, J.~K. Vizzotto, E.~K. Piveta, and A.~R. Du~Bois, ``A monadic semantics for quantum computing in an object oriented language,'' \emph{Science of Computer Programming}, vol. 173, pp. 37--55, 2019.

\bibitem{Chong-EtAl:2017}
F.~T. Chong, D.~Franklin, and M.~Martonosi, ``Programming languages and compiler design for realistic quantum hardware,'' \emph{Nature}, vol. 549, no. 7671, pp. 180--187, 2017.

\bibitem{khan2022embracing}
A.~A. Khan, M.~Fahmideh, A.~Ahmad, M.~Waseem, M.~Niazi, V.~Lahtinen, and T.~Mikkonen, ``Embracing iterations in quantum software: A vision,'' in \emph{Proceedings of the 1st International Workshop on Quantum Programming for Software Engineering}, 2022, pp. 11--14.

\bibitem{Sepulveda-EtAl:2025}
S.~Sepúlveda, R.~Pérez-Castillo, and M.~Piattini, ``A software product line approach for developing hybrid software systems,'' \emph{Information and Software Technology}, vol. 178, 2025.

\bibitem{Wohlin-EtAl:2012}
C.~Wohlin, P.~Runeson, M.~Höst, M.~C. Ohlsson, B.~Regnell, and A.~Wesslén, \emph{Planning}, 1st~ed.\hskip 1em plus 0.5em minus 0.4em\relax London: Springer, 2012, pp. 43--72.

\bibitem{haghparast2024innovative}
M.~Haghparast, E.~Moguel, J.~Garcia-Alonso, T.~Mikkonen, and J.~M. Murillo, ``Innovative approaches to teaching quantum computer programming and quantum software engineering,'' in \emph{2024 IEEE International Conference on Quantum Computing and Engineering (QCE)}, vol.~2.\hskip 1em plus 0.5em minus 0.4em\relax IEEE, 2024, pp. 251--255.

\end{thebibliography}

\end{document}